\documentclass[conference]{IEEEtran}
\IEEEoverridecommandlockouts
\usepackage{cite}
\usepackage{amsmath,amssymb,amsfonts}
\usepackage{algorithm}
\usepackage{algorithmic}
\usepackage{graphicx}
\usepackage{textcomp}
\usepackage{xcolor}
\def\BibTeX{{\rm B\kern-.05em{\sc i\kern-.025em b}\kern-.08em
    T\kern-.1667em\lower.7ex\hbox{E}\kern-.125emX}}
\begin{document}

\title{Understanding User Topic Preferences across Multiple Social Networks\\
}


\author{\IEEEauthorblockN{Ziqing Zhu\IEEEauthorrefmark{1},  Jiuxin Cao\IEEEauthorrefmark{2}\thanks{Jiuxin Cao is corresponding author.}, Tao Zhou\IEEEauthorrefmark{1}, Huiyu Min\IEEEauthorrefmark{1}, Bo Liu\IEEEauthorrefmark{1}}
\IEEEauthorblockA{\IEEEauthorrefmark{1}\textit{School of Computer Science and Engineering},
\textit{Southeast University}, Nanjing, China\\
Email: zzqxztc@seu.edu.cn, zhoutao@seu.edu.cn, minhuiyu@seu.edu.cn, bliu@seu.edu.cn}
\IEEEauthorblockA{\IEEEauthorrefmark{2}\textit{School of Cyber Science and Engineering},
\textit{Southeast University}, Nanjing, China\\
Email: jx.cao@seu.edu.cn}
}

\maketitle

\begin{abstract}
In recent years, social networks have shown diversity in function and applications. People begin to use multiple online social networks simultaneously for different demands. The ability to uncover a user's latent topic and social network preference is critical for community detection, recommendation, and personalized service across social networks. Unfortunately, most current works focus on the single network, necessitating new technology and models to address this issue. This paper proposes a user preference discovery model on multiple social networks. Firstly, the global and local topic concepts are defined, then a latent semantic topic discovery method is used to obtain global and local topic word distributions, along with user topic and social network preferences. After that, the topic distribution characteristics of different social networks are examined, as well as the reasons why users choose one network over another to create a post. Next, a Gibbs sampling algorithm is adopted to obtain the model parameters. In the experiment, we collect data from Twitter, Instagram, and Tumblr websites to build a dataset of multiple social networks. Finally, we compare our research to previous works, and both qualitative and quantitative evaluation results have demonstrated the effectiveness.
\end{abstract}


\section{Introduction}
People's lifestyles have changed tremendously as a result of the rapid development of the Internet. The emergence of online social networks has ushered in a new information era revolution. Online social networks have reshaped everyone's daily life, according to survey statistics revealing that 69\% \footnote[1]{http://www.pewinternet.org/fact-sheet/social-media/} of adults use at least one social network.

In recent years, existing social networks have been enriching themselves with multi-modal information (text, image, video, etc.), highlighting their characteristics due to the diversity and variability of people's needs for information content. To meet different needs, people might use a variety of online social networks. On Twitter, for example, users can join in discussions about current events; on Instagram, users can submit images of their daily lives and share them with others.

Users who have accounts on various online social networks are referred to as ``overlapping users'' in this study. Overlapping users associate information data from several isolated social networks, allowing for a more complete and detailed analysis of user modeling and behavior patterns based on multiple social networks. Integrating user data from different social networks is critical for a more comprehensive view of user behavior patterns.

Many works have been conducted on multiple social networks, such as
overlapping users identification\cite{zhou2015cross,zhang2014meta},
item recommendation\cite{cao2016joint}, 
influence maximization\cite{li2016influence,zhang2015least},
community detection\cite{philip2015mcd,zhu2019community} 
and so on. By designing data fusion strategies, scholars are able to depict users more deeply and uncover the differences in user characteristics.

\begin{figure}[htb]
\centerline{\includegraphics[width=0.4\textwidth]{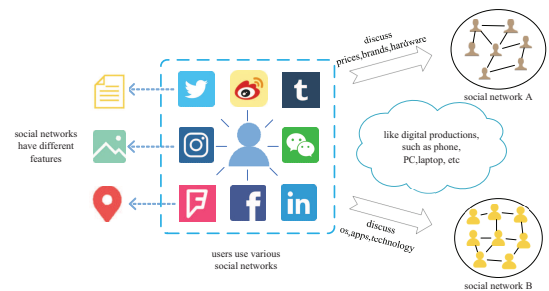}}
\caption{User Topic Preferences across Multiple Social Networks}
\label{fig_multisocialnetwork}
\end{figure}

With the increasing multiple social networks, people are eager to understand why users need to use different social networks and what characteristics attract them. Therefore, this paper aims to investigate users' topic preferences across multiple social networks as a starting point to understand users' motivations. This paper considers users have both consistent and complementary characteristics on various social networks. Consistency refers to the fact that users show similar interests in different social networks. For example, if a user loves digital products, he or she will show his or her love for digital products on different social networks. Complementarity refers to a user presents different aspect characteristics of one topic on various social networks. As shown in Fig.\ref{fig_multisocialnetwork}, if a user likes a digital product, he or she will focus on the brand of the computer or phone, hardware specifications, price, and user experience in network A;
In contrast, he or she will focus on the operating system, related applications, and the technology involved in the digital product in network B. Consistency explores users' similar interests across networks from the common view of the integration across multiple social networks, which is also the basis of the existing research across multiple social networks.

Existing studies\cite{cao2016joint,lee2017analyzing} focus solely on the issue of consistent user topic preferences across multiple social networks. This paper focuses on the complementarity of user topic preferences and analyzes the similarity between user topic preferences and social network topics in different dimensions to explore users' intrinsic motivations for using specific social networks. Social network service providers can collaborate and share data in mutually beneficial ways to provide better social services to users.

The main contributions of this paper are as follows.

1. In this paper, we design a topic preference discovery model across multiple social networks. The model can fuse the data from each social network for unsupervised user topic discovery.

2. This model unifies the two discovery processes of global and local topics corresponding to multiple networks and each specific network, making it possible to explore the similarities and differences of topic word distribution between social networks.

3. Users' topic preferences and social network preferences over topics are also output by this model. As a result, we may analyze each user's characteristics as well as their intrinsic motive for using a specific social network.

In the remainder of this paper, Section 2 reviews the related works; Section 3 describes the research problem; Section 4 gives the multiple social networks topic model; Section 5 evaluates the experiments; Finally, conclusions and future works are provided in Section 5.

\section{Related Works}
\subsection{User Preference Modeling for Single Social Network}
\subsubsection{Based on users' personal data}
Current work on mining users' topic preferences usually uses the data generated by individual data, which include explicitly labeled profile information, such as age, date of birth, gender, interests, and other data. The information also includes users' comments and rating data of various content in social networks\cite{boyd2007social}. However, due to the service limitations of online social networks or personal privacy issues, users' detailed data cannot be easily obtained. Most studies often need to integrate users' implicit feedback data, such as data on users' publicly posted content, forwarded content information, etc. Hong et al.\cite{hong2010empirical} view users' posting data in the Twitter network as documents and use the LDA topic model to model users' topic model. Considering that users' posts in social networks are relatively short, Zhao et al.\cite{zhang2014meta} assume each user's post only has one topic and then restructured a probability generating model based on the LDA model to mine users' topic preferences. 
Cao et al. \cite{cao2017you} combined users' social, interest, and behavioral footprints in the Twitter network and used the matrix decomposition technique to obtain users' topic preference features. Chen et al.\cite{chen2017discerning} divided users' attributes into individual interests and shared interests based on users' item ratings.
Related works have further used multi-modal data to enrich the user's feature content. You et al.\cite{you2016picture} analyzed the images in social networks. They train the image data to extract the hidden features. Then, based on the hidden features, a label propagation algorithm is constructed to achieve images classification. At last, they obtain a user interest model based on image information.
Farnadi et al.\cite{farnadi2018user} used deep learning technique to fuse multi-modal data from social networks. The fused data were classified based on the Big Five personality traits of users from the MyPersonality project, and then mining and analysis of user characteristics was carried out.
\subsubsection{Based on users' friend data}
Mislove et al.\cite{mislove2010you} pointed out that users with similar attributes are more likely to establish friendships and can use the characteristics of users' friends to infer users' attributes.
Chen et al.\cite{chen2016profiling} used relationships and structure characteristics of social networks to construct a user-label attribute prediction model.
Xu et al.\cite{xu2012modeling} modeled users based on three factors, including breaking news, posts from social friends, and user's intrinsic interest.
\subsubsection{Based on users' group data}
Due to the sparsity of users' data, related works have also attempted to integrate data from similar users. Hu et al. \cite{hu2015community} developed a topic model to analyze the change of users' topic preference from a community view; Feng et al. \cite{feng2019user} used user groups to alleviate the sparsity problem of short texts on social media and to detect emotions and relevant topics based on groups; Wang et al.\cite{wang2019community} integrated users' topic and network topologies to catch the topic correlations inside a community.
\subsection{User Preference Modeling for Multiple Social Networks}
%
Considering that users will register accounts and generate data in different social networks, some studies attempt to fuse data of various social networks based on ``overlapping users''. Abel et al.\cite{abel2011analyzing} integrated user information from multiple social networks, such as Flickr, Twitter, and Delicious, and built an attribute tag-based user topic model, which can greatly improve the quality of recommendation algorithms in social networks. Subsequently, they also design to build on a tag-based user model by further integrating the profile description from users' registry information\cite{abel2013cross}. Cao et al.\cite{cao2016joint} used a topic model to simulate users' process of content data generation process in Douan and Weibo, and they correlated the content data of users in different social networks to achieve the mining of user interest preference based on the basic assumption that ``the same user has the same interest preference in different networks''. Lee et al.\cite{lee2017analyzing} used an extended TwitterLDA model to mine users' topic preferences based on their post contents in different social networks and analyzed users' selection preferences for various social networks.

In summary, the study process of user preference modeling, which is gradually transformed from single data in terms of dimensions, types, and features to more complex multi-source heterogeneous data as research objects, the multiple social networks further provide a new research horizon for users' topic preference mining. This paper focuses on analyzing characteristics from a new prospect of multiple social networks compared with the existing works.

\section{Preliminary}


To describe our research work clearly, this paper's main concepts and definitions are given as follows.

\textbf{Definition 1. Multiple Social Networks.} Multiple social networks in this paper are defined as $\boldsymbol {\rm{S}}$. It has the user set $\boldsymbol {\rm{U}}$. Every social network is $s\in\boldsymbol {\rm{S}}$. $\boldsymbol {\rm{S}}_u$ represents social network set of user $u$. $\boldsymbol {\rm{C}}_u^s$ represents content set of user $u$ in social networks $s$.

\textbf{Definition 2. User Topic Preference.} User topic preference represents the topic interest of the user. It is expressed by multinomial distribution over topics, represented as $\theta_u$. The topic $z$ has a probability distribution of belonging to user $u$.

\textbf{Definition 3. Global Topics and Local Topics.} Global topics represent topics that are discovered based on fused multiple social networks. Local Topics represent topics that are discovered based on a specific single social network. In this paper, the global topic distribution of users is described as a multinomial distribution $\varphi_z^p$ which is a probability distribution of vocabulary over topic across multiple social networks. Local topic distribution of users is described as a multinomial distribution $\varphi_z^s$, a probability distribution of vocabulary over the topic in social network $s$. Each word $w$ has a probability of belonging to topic $z$. Every post $c$ in content set $\boldsymbol {\rm{C}}_u^s$ can generate word set $\boldsymbol {\rm{W}}_{u,c}$.

\textbf{Definition 4. User Social Network Preference.} This paper describes user social network preference as a multinomial distribution $\rho_{uz} $ which is a probability distribution of choosing social network by user $u$ over topic $z$. The social network $s$ has the probability of belonging to $z$.

This paper aims to obtain users' topic preference features based on their content data and preference data of social networks for multiple social networks, analyze users' topic contents, and explore the similarity and difference of topic word distribution in different social networks. The formalization of the problem is as follows. Given multiple social network set $\boldsymbol {\rm{S}}$. The model will output user topic preference features $\theta_u$, user global topic word distribution $\varphi_z^p$ and user local topic word distribution $\varphi_z^s$, and social network selection preference $\rho_{uz}$. The relevant notations in this paper are illustrated in Tab.\ref{tab_Descriptions of symbol}.
\begin{table}
\caption{Descriptions of symbol}
\newcommand{\tabincell}[2]{\begin{tabular}{@{}#1@{}}#2\end{tabular}}
\begin{center}
\begin{tabular}{  c| c}
\hline
\textbf{Symbol} & \textbf{Description}\\
\hline
$\boldsymbol {\rm{S}}$, $\boldsymbol {\rm{U}}$, $\boldsymbol {\rm{V}}$& \tabincell{c}{social network set, user set,vocabulary set}\\
\hline
$\boldsymbol {\rm{C}}_u^s$, $\boldsymbol {\rm{W}}_{u,c}$& \tabincell{c}{user $u$'s content set in social network $s$,\\word set of user $u$'s post $c$}\\
\hline
$z$, $c$, $w$& topic, post, word\\
\hline
$x$& switching variable\\
\hline
$\theta_u$& topic distribution of user $u$\\
\hline
$\phi_z^p$, $\phi_z^s$ & \tabincell{c}{word distribution over the global topic $z$,\\word distribution over the local topic $z$}\\
\hline
$\phi_z^p$& word distribution over the background topic\\
\hline
$\rho_{uz}$& social network distribution of user $u$ over topic $z$\\
\hline
$\sigma_{uz}$& \tabincell{c}{switching variable distribution to control whether \\ topic word is generated from $\phi_z^p$ or $\phi_z^s$}\\
\hline
$\sigma_B$& \tabincell{c}{switching variable distribution to control whether \\ or not word is generated from $\phi_B$}\\
\hline
$\boldsymbol {\alpha$, $\beta^p$, $\beta^s$, $\beta_B$, $\lambda$, $\tau}$ & dirichlet prior hyperparameters\\
\hline
\end{tabular}
\label{tab_Descriptions of symbol}
\end{center}
\end{table}
\section{User Topic Preference Model}
This section will describe the model structure and give the algorithm for solving the model parameters.
\subsection{Model Structure}
In social networks, the aggregation of high-frequency co-occurring words often represents a topic. This paper needs to aggregate high-frequency co-occurring words from the global perspective across multiple social networks and find out the consistent global topic content. Furthermore, the words that have a co-occurrence relationship with a global topic in a specific social network are aggregated as local topics, showing different specific topic words. The aggregation of high-frequency co-occurring words prompts us to use the LDA-based topic model to mine user topic preferences in multiple social network situations. The LDA model is an unsupervised cluster that uses co-occurring words to generate a topic.

\begin{figure}[htb]
\centerline{\includegraphics[width=0.35\textwidth]{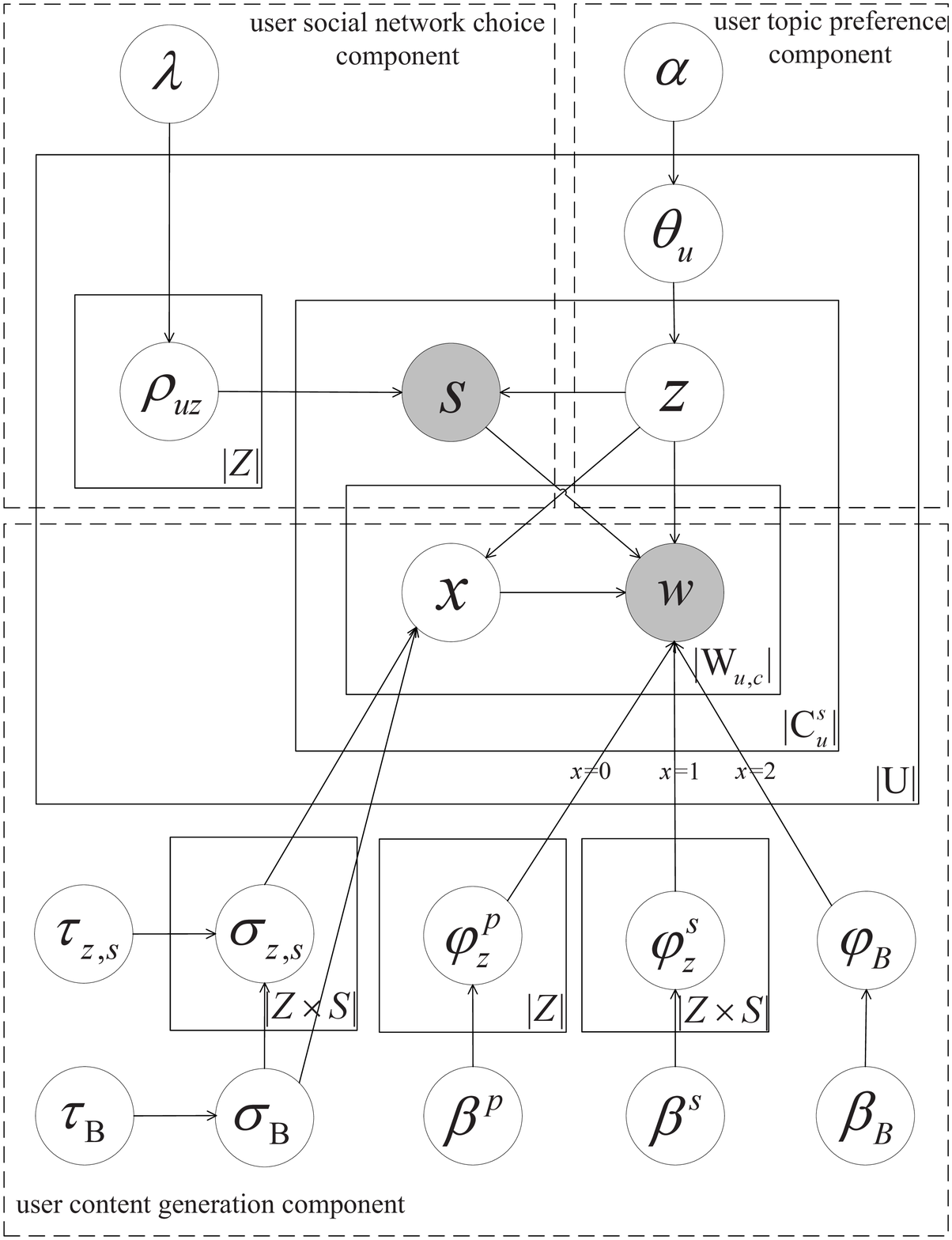}}
\caption{The Graphical Representation of Proposed Model}
\label{fig_modelstructure}
\end{figure}

We design a Multiple Social Networks Topic Model (MSNT). The basic structure of the model is shown in Fig.\ref{fig_modelstructure}, which mainly contains the user topic preference component, user content generation component, and user social network choice component.
\subsubsection{\textbf{User topic preference component}}
Users have kinds of topic content in social networks, and each user has unique topic preferences than others. To model this situation, we view users' topic preference distribution as a multinomial distribution over topics $\theta_u$. Each topic $z$ can be generated from the distribution.
\subsubsection{\textbf{User content generation component}}
Users create posts based on their topic preferences. In our MSNT model, the word generation process over the topic is divided into two types of processes.

\emph{Global topics words generation:} In multiple social network situations, the model first uses a multinomial distribution $\varphi_z^p$ to generate word $w$ over topic $z$ based on the user's all content data. This can find co-occurring words from the global perspective on all social networks. We can discover users' topics based on complete multiple social network data to reflect the intrinsic preferences of users.

\emph{Local topics words generation:} The specific social network may also influence a user to use different words to make posts. The model uses a multinomial distribution $\varphi_z^s$ to generate word $w$ over topic $z$ only on user content data in a specific social network. We can find topics to reflect the specific content characteristic based on the single social network.

In order to unify users' global topics and local topics, this paper describes a switch variable $x$ to control the selection of words for topic content. The probability distribution that $x$ takes the 0 or 1 is denoted as a binomial distribution $\sigma_{zs}$. The binomial distribution can represent the latent proportion distribution of global topic words and local topic words. When $x=0$, the users select global topic word $w$ under topic $z$, and the distribution of this process is denoted as multinomial distribution $\varphi_z^p$. When $x=1$, the users generate local topic word $w$ with topic $z$ in social network $s$, and distribution of this process is denoted as multinomial distribution $\varphi_z^s$. The semantic content between the global topic and local topic is associated with each other by switch variables. In addition, we assume that a background topic captures the background words used across the multiple social networks to improve the ability to mining other topics. Therefore, the switch variable $x$ first takes the value tag from a binomial distribution $\sigma_B$. If $x$ get $2$, word $w$ is generated from a background topic word multinomial distribution $\varphi_B$. If $x$ is not $2$, the switch variable $x$ then takes value from the binomial distribution $\sigma_{zs}$. The word $w$ is from the global topic or local topic depending on the variable $x$ value of $0$ or $1$.

This paper also assumes each post $c$ has only one topic $z$ by referring to the work on social network topic mining\cite{zhao2011comparing,lee2017analyzing}, because the length of posts in social networks is much smaller than the length of normal documents, and the user will only focus on one topic when creating a post.
\subsubsection{\textbf{User social network choice component}}
The service differences in social networks drive users to choose different social networks. Users may choose different social networks to post content based on topics. Users' social network choice is influenced by potential relevance between users' topic preferences and social network topics. 
In the model, the probability distribution that user $u$ chooses social network $s$ with topic $z$ is described as a multinomial distribution $\rho_{uz}$.

\subsubsection{whole generative process} We summarize the above generative process as follows.
\begin{enumerate}[\IEEEsetlabelwidth{8)}]
\item Sample the background topic vocabulary Multinomial distribution $\varphi_B$ from a prior Dirichlet distribution: $\varphi_B \sim Dir(\beta_B) $; Sampe the Binomial distribution of whether user's content word is background word from a prior beta distribution: $\sigma_{B} \sim Dir(\tau_{B}) $;
\item For each topic $z$,
\begin{enumerate}[\IEEEsetlabelwidth{8)}]
\item Sample the global topic vocabulary Multinomial distribution $\varphi_z^p$ from a Dirichlet prior distribution. $\varphi_z^p \sim Dir(\beta^p) $;
\item For each social network $s$,
\begin{enumerate}[\IEEEsetlabelwidth{8)}]
\item Sample the local topic vocabulary Multinomial distribution in social networks $s$ from a prior Dirichlet distribution: $\varphi_z^s \sim Dir(\beta^s)$;
    Sample the Binomial distribution of whether user's content word is global or local topic word from a prior beta distribution: $\sigma_{zs} \sim Dir(\tau_{zs}) $;
\item Sample the user social network preference Multinomial distribution from a prior Dirichlet distribution: $\rho_{uz} \sim Dir(\lambda_{uz})$;
\end{enumerate}
\end{enumerate}
\item For each user $u$,
\begin{enumerate}[\IEEEsetlabelwidth{8)}]
\item Sample the user topic preference Multinomial distribution from a prior Dirichlet distribution: $\theta_u \sim Dir(\alpha)$.
\item For each user $u$'s post,
\begin{enumerate}[\IEEEsetlabelwidth{8)}]
\item Sample a topic assignment $z$ from Multinomial distribution
: $z \sim Mult(\theta_u)$,
\item Sample a social network indicator $s$ from Multinomial distribution
: $s \sim Mult(\rho_{uz})$;
\item For each word $w$,
\begin{enumerate}[\IEEEsetlabelwidth{8)}]
\item Sample a $x$ value from binomial distribution
: $x \sim Bern(\sigma_{B})$,
\item If $x=2$, sample a word $w$ from Multinomial distribution
: $w \sim Mult(\varphi_B)$;
\item If $x \neq 2$, sample a $x$ value from binomial distribution
: $x \sim Bern(\sigma_{zs})$. If $x =0$, sample a word $w$ from Multinomial distribution
: $ w \sim Mult(\varphi_z^p)$. If $x =1$, sample a word $w$ from Multinomial distribution
: $w \sim Mult(\varphi_z^s)$.
\end{enumerate}
\end{enumerate}
\end{enumerate}
\end{enumerate}
\subsection{Model Inference}
For the brevity of description, this paper lets hyper-parameters
$\{\alpha, \beta^p,\beta^s, \beta_B, \tau, \lambda\}$ be denoted as $\boldsymbol{\Phi}$
, and hidden variables $\{\theta, \varphi^p_{uz},$ $ \varphi^s_z, \phi_B,$ $ \sigma_{zs}, \sigma_B, \rho_{uz} \}$ as $\boldsymbol{\Omega}$.
The latent variables conditioned on the observed variables, namely $p(\boldsymbol{z},\boldsymbol{x}|\boldsymbol{w},\boldsymbol{u},$ $\boldsymbol{s},\boldsymbol{\Phi})$ are needed to estimate. This paper uses the collapsed Gibbs sampling method to obtain hidden variables $\boldsymbol{\Omega}$ in our model. In the Gibbs sampling algorithm, the new value of the parameter is determined by the previous parameter value of the model, and the model parameters are constantly updated iteratively to finally complete the solution of the model parameters.

\textbf{Sampling topic $z$ of the post.} The length of a post in social network is generally much smaller than the length of a normal document. In most cases, users' posts only describe one topic. The sampling formula for the topic of the post is Eq.(\ref{eq_sampling_z}).
\begin{equation}\label{eq_sampling_z}
\begin{aligned}
&p({z_{c_i}}\!=\! k|\boldsymbol{z}_{\neg c_i},\boldsymbol{x},\boldsymbol{w},\boldsymbol{u},\boldsymbol{s},\boldsymbol{\Phi})=\frac{p({z_{c_i}} \!=\! k, \boldsymbol{z}_{\neg c_i},\boldsymbol{x},\boldsymbol{w},\boldsymbol{u},\boldsymbol{s},\boldsymbol{\Phi})}{p(\boldsymbol{z}_{\neg c_i},\boldsymbol{x},\boldsymbol{w},\boldsymbol{u},\boldsymbol{s},\boldsymbol{\Phi})} \\ &\propto \frac{n_{{u, \neg c_i}}^{(k)} + {\alpha _k}}{\sum _{k \in \boldsymbol{Z}} (n_{{u, \neg c_i}}^{(k)}+{\alpha _k}) } \cdot \frac{n_{u,\neg c_i}^{( h )} + {\lambda _h}}{\sum _{h \in \boldsymbol{S} } ( n_{u,\neg c_i}^{( h )} + {\lambda _h})} \cdot
\prod \limits_{x_w \neq 2, w \in c_i} \\
& \left\{
 \begin{aligned}
      \frac{n_{{k, \neg c_i}}^{(w)} + {\beta^p_k}}{\sum _{w \in \boldsymbol{V}} (n_{{k, \neg c_i}}^{(w)}+{\beta^p_k}) } \cdot \frac{n_{{h,k, \neg c_i}}^{(0)} + {\tau _{zs,0}}}{\sum_{x \in (0,1)} n_{{h,k, \neg c_i}}^{(x)}+{\tau _{zs,x}}}, x_w=0\\
      \frac{n_{{h, k, \neg c_i}}^{(w)} + {\beta^h_k}}{\sum _{w \in \boldsymbol{V}} (n_{{h,k, \neg c_i}}^{(w)}+{\beta^h_k}) } \cdot \frac{n_{{h,k, \neg c_i}}^{(1)} + {\tau _{zs,1}}}{\sum_{x \in (0,1)} n_{{h,k, \neg c_i}}^{(x)}+{\tau _{zs,x}}}, x_w=1
  \end{aligned}
  \right.
\end{aligned}
\end{equation}
where $x_w \neq 2, w \in c_i$ denotes the words which are not background topic words. $n_{{u, \neg c_i}}^{(k)}$ is the number of posts of user $u$ which are associated with topic $k$, excluding the current post instance $c_i$. $n_{u,\neg c_i}^{( h )}$ is the number of social network $h$ of user $u$'s posts according to topic $k$, excluding the current post instance $c_i$. $n_{{k, \neg c_i}}^{(w)}$ is the number of words according to topic $k$, excluding the current post instance $c_i$. $n_{{h, k, \neg c_i}}^{(w)}$ is the number of words according to topic $k$ in social network $h$, excluding the current post instance $c_i$. $n_{{h,k, \neg c_i}}^{(0)}$ is the number of global topic words according to topic $k$ in social network $h$. $n_{{h,k, \neg c_i}}^{(1)}$ is the number of local topic words according to topic $k$ in social network $h$.

\textbf{Sampling switch variable $x$.} As described above, the value of $x$ is the process of users' choice of words in the multiple social networks. The sampling formula of variable $x$ is Eq.(\ref{eq_sampling_x_0}), (\ref{eq_sampling_x_1}) or (\ref{eq_sampling_x_2}).
\begin{equation}\label{eq_sampling_x_0}
\begin{aligned}
&p({x_{w_i}}\!=\! 0|\boldsymbol{x}_{\neg w_i},\boldsymbol{z},\boldsymbol{w},\boldsymbol{u},\boldsymbol{s},\boldsymbol{\Phi})\!=\!\frac{p({x_{w_i}} \!=\! 0, \boldsymbol{x}_{\neg w_i},\boldsymbol{z},\boldsymbol{w},\boldsymbol{u},\boldsymbol{s},\boldsymbol{\Phi})}{p(\boldsymbol{x}_{\neg w_i},\boldsymbol{z},\boldsymbol{w},\boldsymbol{u},\boldsymbol{s},\boldsymbol{\Phi})} \\
& \!\propto\! \frac{n_{{B, \neg w_i}}^{(0+1)} + {\tau_{B,0+1}}}{n_{{B, \neg w_i}}^{(2)}+n_{{B, \neg w_i}}^{(0+1)}+ {\tau_{B,2}} + {\tau_{B,0+1}}} \!\cdot\! \frac{n_{{h,k, \neg c_i}}^{(0)} + {\tau _{zs,0}}}{\sum_{x \in (0,1)} n_{{h,k, \neg w_i}}^{(x)}+{\tau _{zs,x}}} \\
& \qquad \qquad \qquad \qquad \qquad \!\cdot\! \frac{n_{{k, \neg w_i}}^{(w_i)} + {\beta^p_k}}{\sum _{w \in \boldsymbol{V}} (n_{{k, \neg w_i}}^{(w)}+{\beta^p_k})}\\
\end{aligned}
\end{equation}
\begin{equation}\label{eq_sampling_x_1}
\begin{aligned}
&p({x_{w_i}}\!=\! 1|\boldsymbol{x}_{\neg w_i},\boldsymbol{z},\boldsymbol{w},\boldsymbol{u},\boldsymbol{s},\boldsymbol{\Phi})\!=\!\frac{p({x_{w_i}} \!=\! 1, \boldsymbol{x}_{\neg w_i},\boldsymbol{z},\boldsymbol{w},\boldsymbol{u},\boldsymbol{s},\boldsymbol{\Phi})}{p(\boldsymbol{x}_{\neg w_i},\boldsymbol{z},\boldsymbol{w},\boldsymbol{u},\boldsymbol{s},\boldsymbol{\Phi})} \\
& \!\propto\! \frac{n_{{B, \neg w_i}}^{(0+1)} + {\tau_{B,0+1}}}{n_{{B, \neg w_i}}^{(2)}+n_{{B, \neg w_i}}^{(0+1)}+ {\tau_{B,2}} + {\tau_{B,0+1}}} \!\cdot\! \frac{n_{{h,k, \neg w_i}}^{(1)} + {\tau _{zs,1}}}{\sum_{x \in (0,1)} n_{{h,k, \neg w_i}}^{(x)}+{\tau _{zs,x}}} \\
& \qquad \qquad \qquad \qquad \qquad \!\cdot\!  \frac{n_{{h, k, \neg w_i}}^{(w_i)} + {\beta^s_k}}{\sum _{w \in \boldsymbol{V}} (n_{{h,k, \neg w_i}}^{(w)}+{\beta^s_k}) }\\
\end{aligned}
\end{equation}
\begin{equation}\label{eq_sampling_x_2}
\begin{aligned}
&p({x_{w_i}}\!=\! 2|\boldsymbol{x}_{\neg w_i},\boldsymbol{z},\boldsymbol{w},\boldsymbol{u},\boldsymbol{s},\boldsymbol{\Phi})\!=\!\frac{p({x_{w_i}} \!=\! 2, \boldsymbol{x}_{\neg w_i},\boldsymbol{z},\boldsymbol{w},\boldsymbol{u},\boldsymbol{s},\boldsymbol{\Phi})}{p(\boldsymbol{x}_{\neg w_i},\boldsymbol{z},\boldsymbol{w},\boldsymbol{u},\boldsymbol{s},\boldsymbol{\Phi})} \\
& \!\propto\! \frac{n_{{B, \neg w_i}}^{(2)} +{\tau_{B,2}}}{n_{{B, \neg w_i}}^{(2)}+n_{{B, \neg w_i}}^{(0+1)}+ {\tau_{B,2}} + {\tau_{B,0+1}}}\! \cdot\! \frac{n_{{B, \neg w_i}}^{(w_i)} + {\beta_B}}{\sum _{w \in \boldsymbol{V}} (n_{{B, \neg w_i}}^{(w)}+{\beta_B}) }
\end{aligned}
\end{equation}
where $n_{{B, \neg w_i}}^{(0+1)}$ is the number of words which are not associated with background topic, excluding the current word instance $w_i$. $n_{{B, \neg w_i}}^{(2)}$ is the number of words which are associated with background topic, excluding the current word instance $w_i$. $n_{{h,k, \neg w_i}}^{(0)}$ is the number of global topic words according to topic $k$ in social network $h$, excluding the current word instance $w_i$. $n_{{h,k, \neg w_i}}^{(1)}$ is the number of local topic words according to topic $k$ in social network $h$, excluding the current word instance $w_i$. $n_{{k, \neg w_i}}^{(w)}$ is the number of words according to topic $k$, excluding the current word instance $w_i$. $n_{{h, k, \neg w_i}}^{(w)}$ is the number of words according to topic $k$ in social network $h$, excluding the current word instance $w_i$. $n_{{B, \neg w_i}}^{(w)}$ is the number of background topic words, excluding the current word instance $w_i$.

After Gibbs algorithm sampling, the model parameters are calculated as in Eq.(\ref{eq_theta}) to Eq.(\ref{eq_rho}).
\begin{equation}\label{eq_theta}
\theta_u=\frac{n_u^{(k)} + {\alpha _k}}{\sum _{k \in \boldsymbol{Z}} (n_u^{(k)}+{\alpha _k}) }
\end{equation}
\begin{equation}\label{eq_phi_p}
\varphi_z^p=\frac{n_z^{(w)} + {\beta_z^p}}{\sum _{w \in \boldsymbol{V}} (n_z^{(w)}+{\beta_z^p}) }
\end{equation}
\begin{equation}\label{eq_phi_s}
\varphi_z^s=\frac{n_{s,z}^{(w)} + {\beta_z^s}}{\sum _{w \in \boldsymbol{V}} (n_{s,z}^{(w)}+{\beta_z^s}) }
\end{equation}
\begin{equation}\label{eq_rho}
\rho_{uz}=\frac{n_{u,z}^{(h)} + {\lambda_z}}{\sum _{h \in \boldsymbol{S}} (n_{u,z}^{(h)}+{\lambda_z}) }
\end{equation}

The algorithm of parameters inference for MSNT is described as Algorithm \ref{alg_Inference_for_MSNT}.
\begin{figure}[htb]
\begin{algorithm}[H]
\caption{Inference for MSNT}
\label{alg_Inference_for_MSNT}
\renewcommand{\algorithmicrequire}{\textbf{Input:}}
\renewcommand{\algorithmicensure}{\textbf{Output:}}
\begin{algorithmic}[1]
\REQUIRE $\boldsymbol {\rm{S}}$
\ENSURE $\boldsymbol{\Omega}$
\STATE Initialize $\boldsymbol{\Phi}$, $\boldsymbol{\Omega}$
\FOR{ iter=1 to max\_iter\_number }
 \FOR{each user $u \in \boldsymbol {\rm{U}}$}
  \FOR{each social network $s \in \boldsymbol {\rm S}_u$}
   \FOR{each post $c \in \boldsymbol {\rm C}^s_u$}
    \FOR{each word $w \in \boldsymbol {\rm W}^s_c$}
    \STATE Sample word state $x$ according to Eq.(\ref{eq_sampling_x_0}), Eq.(\ref{eq_sampling_x_1}) or Eq.(\ref{eq_sampling_x_2})
    \STATE Update word state $x$
    \ENDFOR
    \STATE Sample post topic $z$ according to Eq.(\ref{eq_sampling_z})
   \ENDFOR
  \ENDFOR
 \ENDFOR
 \STATE update $\boldsymbol{\Omega}$
\ENDFOR
\end{algorithmic}
\end{algorithm}
\end{figure}

Line 2 in the algorithm, max\_iter\_number indicates the maximum number of iterations. The termination condition can also be replaced by other convergence conditions. The time complexity of the algorithm is $O(|\boldsymbol {\rm{U}}|\times |\boldsymbol {\rm S}_u | \times |\boldsymbol {\rm C}^s_u| \times |\boldsymbol {\rm W}^s_c|)$.

\section{Experiment}
This section performs some experiments to verify the model effectiveness of mining user topic preference. Specifically, the dataset and experiment setup are first given. Then, the evaluation criteria are described in detail. We show the evaluation of the proposed model MSNT compared with other works. Finally, the analysis and discussion of the results will be presented.
\subsection{Dataset}
This paper uses a dataset of multiple social networks, including Twitter, Instagram, and Tumblr. Twitter is a text-based short-text social network. Instagram is a photo-sharing social network. Tumblr is a multimedia blog-based social network that supports a wide range of rich media such as pictures, videos, etc.

The user account information in the dataset is crawled from the business card website About.me, from which to link multiple online identities. The total number of users is 5747. The user accounts include 2076 Twitter users, 2014 Instagram users, and 1655 Tumblr users.
Based on these users, user content data are crawled from the connected social networks respectively. The users' content data spans from January 2017 to January 2018, where the non-text type data, such as images, are converted into word tags by Google's deep learning-based image annotation model\footnote[2]{https://www.googlevision.com}.

The statistics of the dataset are shown in Tab.\ref{tab_numble_user} and Tab.\ref{tab_number_content}. Each element in Tab.\ref{tab_numble_user} is the number of identical users corresponding to two social networks.In addition, there are 935 identical users who link all three networks. We conduct topic mining based on their content data. The number of users' content data is shown in Tab.\ref{tab_number_content}.
\begin{table}[htb]
\caption{The Number of Users Across Social Networks}
\begin{center}
\begin{tabular}{|c|c|c|c|}
\hline
&\textbf{Twitter}& \textbf{Instagram}& \textbf{Tumblr} \\
\hline
\textbf{Twitter}&2076& 1435 &1076 \\
\hline
\textbf{Instagram}&-& 2014 & 1014 \\
\hline
\textbf{Tumblr}&-&- & 1665 \\
\hline
\end{tabular}
\label{tab_numble_user}
\end{center}
\end{table}
\begin{table}[htb]
\caption{The Number of Contents}
\begin{center}
\begin{tabular}{|c|c|c|c|}
\hline
&\textbf{Twitter}& \textbf{Instagram}& \textbf{Tumblr} \\
\hline
\textbf{content}&1,257,055 & 302,192 &413,359 \\
\hline
\end{tabular}
\label{tab_number_content}
\end{center}
\end{table}
\subsection{Setup}
The number of topics $k$ is determined by experiment performance. About other prior parameters, referring to related research works\cite{lee2017analyzing,qian2016multi,heinrich2005parameter}, the prior hyperparameters are taken as an empirical value, $\alpha=50/k$, $\beta^p=\beta^s=\beta_B=\lambda_{zs}=\lambda_z=0.01$.
\subsection{Topic Content Quality}
In this paper, we first compare and analyze the model's topic discovery quality.
\subsubsection{Evaluation metrics}
Viewing users as a document set, the model in this paper can be regarded as a document topic discovery model. Therefore, this paper uses the Perplexity\cite{heinrich2005parameter}, Likelihood\cite{heinrich2005parameter} and word Pointwise Mutual Information Score (PMI-score)\cite{cheng2014btm} metrics to evaluate the model performance.

Specifically, the perplexity is a widely used evaluation metric in the field of natural language processing, and the higher quality of the model is corresponding to the lower perplexity value. Its specific form is shown in Eq.(\ref{eq_perplexity}).
\begin{equation}\label{eq_perplexity}
Perplexity(W|M)=exp(-\frac{\sum _{d=1}^D logP(w_d|M)}{\sum _{d=1}^D N_d })
\end{equation}
where $P(w_d|M  )$ denotes the probability that document $w_d$ is output by model $M$ and $N_d$ denotes the length of document $w_d$. Here, since this paper considers each post of user as a document and marks it as one topic. Then $P(w_d|M)=\sum_{k \in Z}\theta^k_u \prod \limits_{w \in W_d} P(w_d|M )$.

The likelihood calculates the probability of generating the sentence in the test corpus. The likelihood value is higher, and it can show better quality of the model. Its specific form is shown in Eq.(\ref{eq_likelihood}).
\begin{equation}\label{eq_likelihood}
Likelihood(W|M)=\sum _{d=1}^D logP(w_d|M)
\end{equation}
Generally, the multiplication value of probability is transformed into summation by log operation to prevent the multiplication value from being too small.

The PMI-score measures the topic quality based on the co-occurrence word pairs in one topic. The words within the topic are correlated, and the higher mutual information score indicates a higher model quality. Its specific form is shown in Eq.(\ref{eq_PMI_score}).
\begin{equation}\label{eq_PMI_score}
PMIScore=\frac{1}{K}\sum _{k \in Z}\frac{1}{T(T-1)}\sum_{1\leq i<j\leq T} PMI_k (w_i,w_j)
\end{equation}
where, $K$ denotes the number of topic. $T$ denotes the number of words in topic. The word mutual information calculates the score of the $Top-T$ words with the highest probability of occurrence in the topic. $PMI_k(w_i,w_j)=log\frac{p(w_i,w_j)}{p(w_i)p(w_j)}$. $p(w_i,w_j)$ is probability of co-occurring word pair $w_i$ and $w_j$. $p(w)$ is probability of occurrence of word $w_i$.
\subsubsection{Baselines}
To demonstrate the effectiveness of the proposed MSNT model, we choose different representative baseline methods as follows.

\textbf{LDA}\cite{heinrich2005parameter}, which is a classical topic discovery model. During topic discovery, the model integrates multiple social network data directly without distinguishing the source of the data, and one document consists of all contents generated by the user.

\textbf{TwitterLDA}\cite{zhao2011comparing}, which takes into account the short-text features in social networks, links one post to only one topic in the topic discovery process, and the model integrates data of multiple social networks without distinguishing the sources of data.

\textbf{MTM}\cite{qian2016multi}, the model can mine topic-related opinion words from different corpus. During the topic discovery process, the model can distinguish the source corpus of the data. In the experiments, various social networks are corresponding to the different corpus. The topic discovery part of the model is also equivalent to the ccLDA [35] topic discovery model.

\textbf{MutilLDA}\cite{lee2017analyzing}, which has similar features to TwitterLDA, links one post to only one topic in the topic discovery process. The model considers users' choice preferences of social networks across multiple social networks.

\textbf{MSNT}, which is the topic discovery model across multiple social networks in this paper.
\subsubsection{Results Analysis}
\begin{figure}[htb]
\centerline{\includegraphics[width=0.38\textwidth]{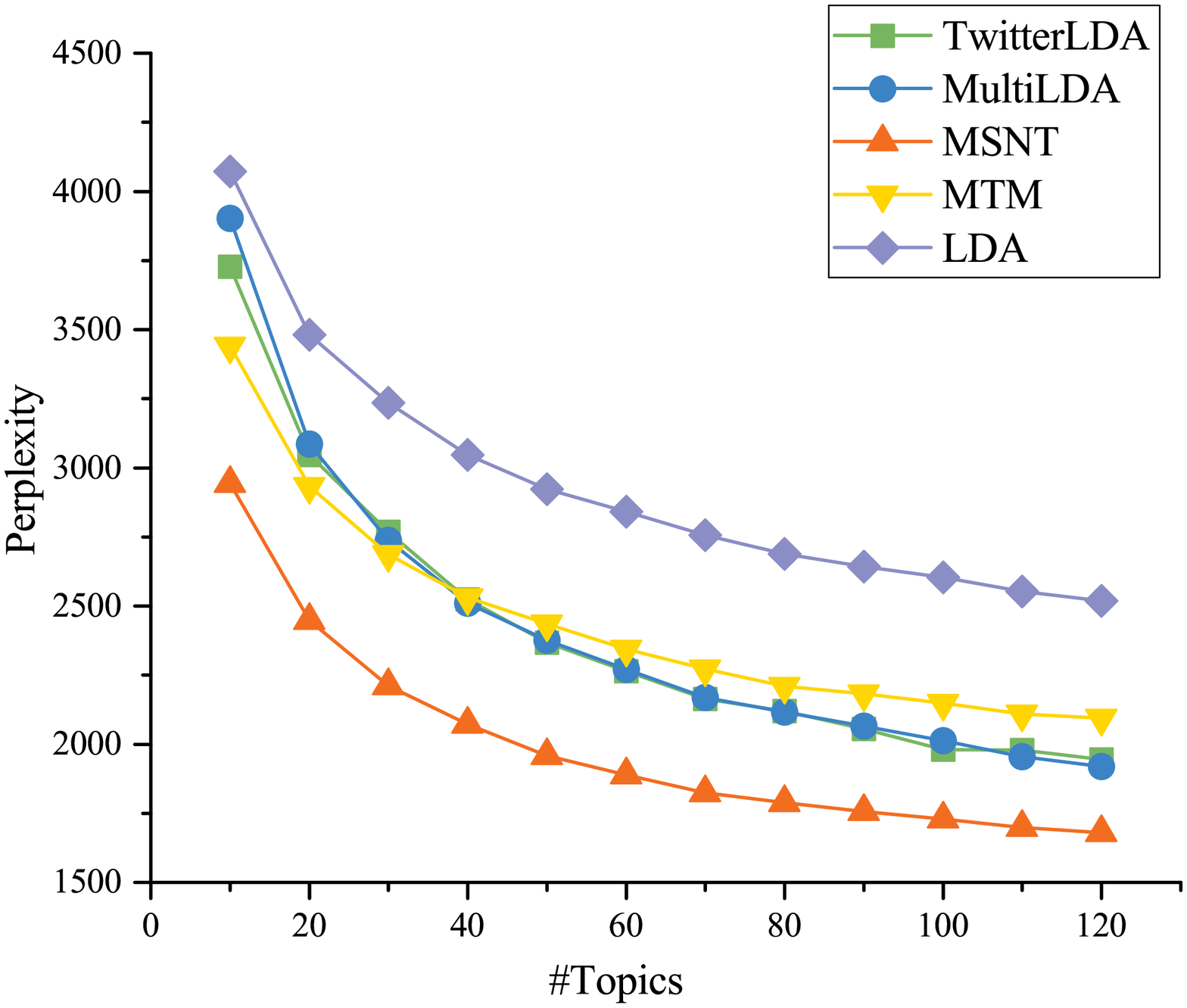}}
\caption{Perplexity Comparison}
\label{fig_perplexityresults}
\end{figure}

First, we show perplexity experiment results. The lower perplexity value is better. It can be observed that perplexity values of all models tends to decrease as the number of topic increases shown in Fig.\ref{fig_perplexityresults}. 

Starting from the number of topics, about 80, the decrease of the perplexity value gradually tends to be smooth. In particular, the perplexity values of TwitterLDA, MultiLDA, and the MSNT are significantly lower than MTM and LDA. The reason is that TwitterLDA, MultiLDA, and MSNT are specific to short texts of social networks, and the background topics in these models reduce the influence of noisy words in social networks. The perplexity experiment results of MSNT in this paper are better than TwitterLDA and MultiLDA, because MSNT is designed to apply to the scenario of multiple social networks compared with TwitterLDA. For MultiLDA, MSNT involves users' preference of choosing social networks and mining users' topic preference from a global view on fusing multiple social networks data. It can also discover differences in local topic content across multiple social networks. Overall, MSNT obtains better performance than other works due to modeling global topics and local topics being more reasonable in multiple social network scenarios, i.e., consistency and complementarity co-exist.

\begin{figure}[htb]
\centerline{\includegraphics[width=0.38\textwidth]{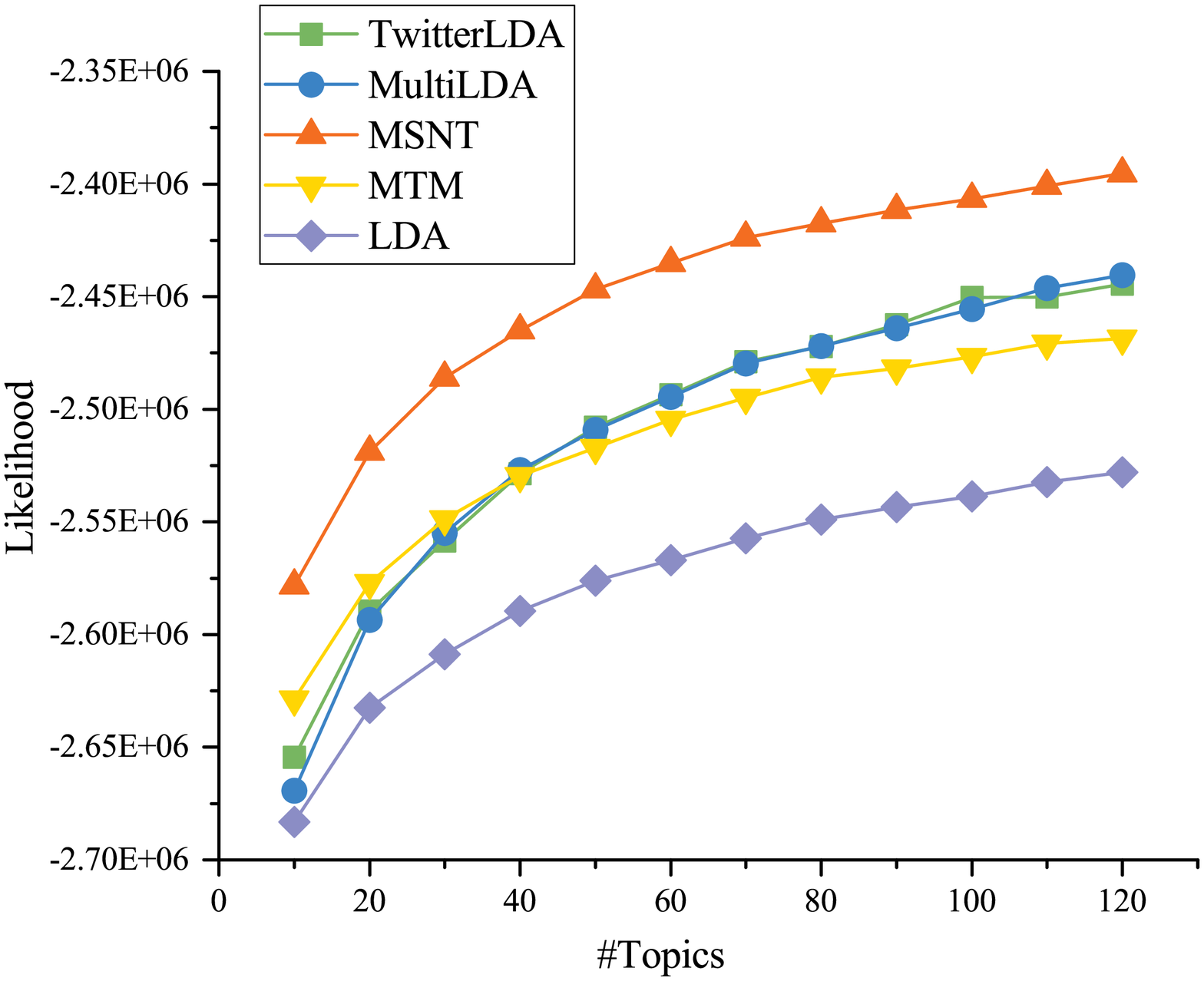}}
\caption{Likelihood Comparison}
\label{fig_likehoodresults}
\end{figure}

The likelihood value can describe the generation probability of a sentence in the test set based on the model. Its rising and decreasing trends are similar to perplexity.
From experiment results shown in Fig.\ref{fig_likehoodresults}, 
we can also see that the increasing trend of likelihood gradually becomes smooth as the number of topics increases. Compared with other works, MSNT significantly gets better experiment results.

\begin{figure}[htb]
\centerline{\includegraphics[width=0.38\textwidth]{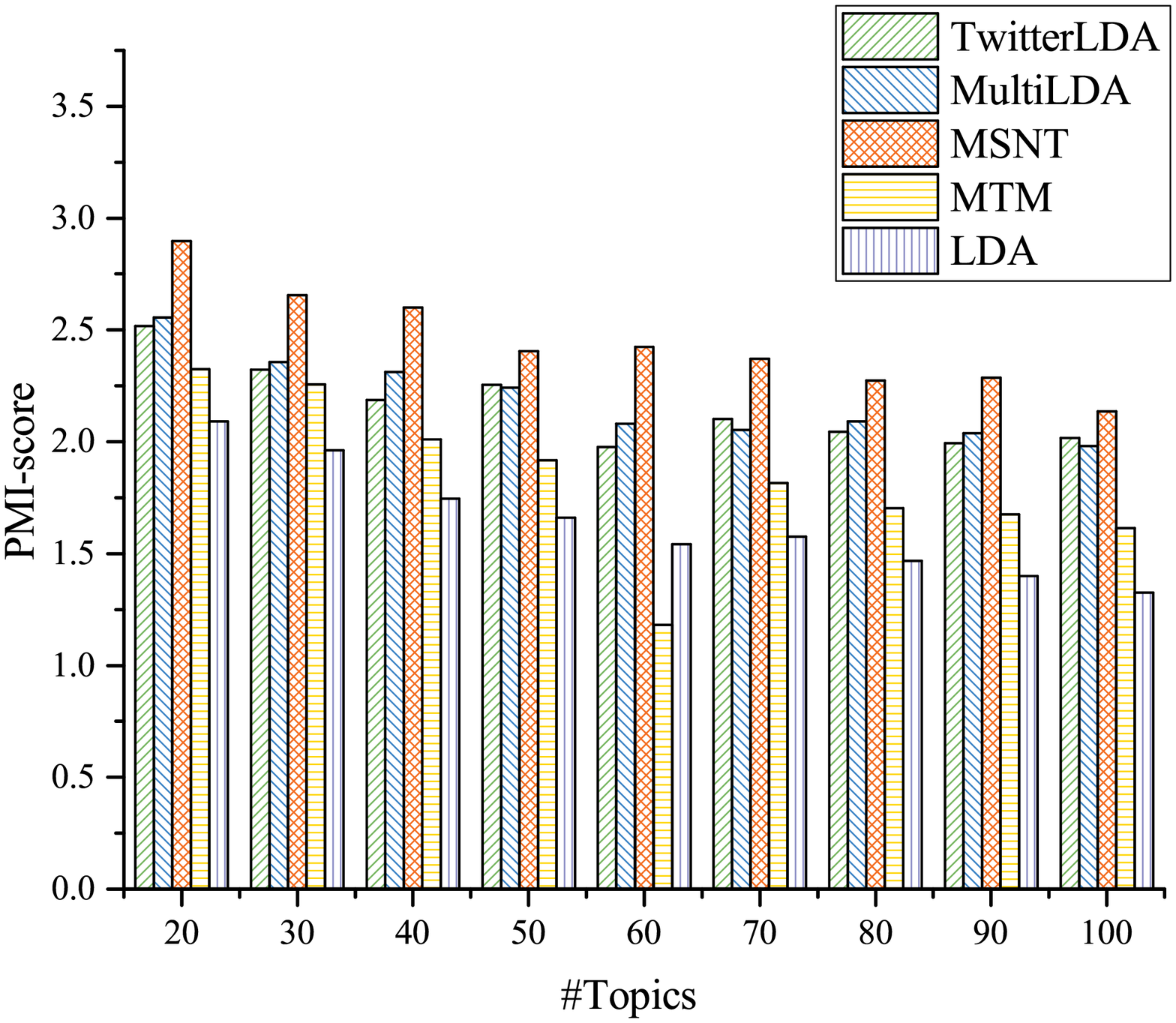}}
\caption{PMI-scroe Comparison}
\label{fig_PMIscroeresults}
\end{figure}

For word PMI-score, which is used to measure the lexical relevance in one topic, the higher mutual information score indicates a higher model quality. Experiment results are shown in Fig.\ref{fig_PMIscroeresults}. 
It can be seen that the number of topic is from 20 to 100, and the top 50 words are taken in one topic to calculate the word PMI-score under different topic numbers.

The figure shows that TwitterLDA, MultiLDA, and MSTN perform better than MTM and LDA because these models are designed based on social networks with short texts. The MSTN in this paper achieves better results than TwitterLDA and MultiLDA because MSTN considers local topics that are more likely to have some feature topic words in a specific social network. These words are easier to form co-occurring word pairs in a specific social network, thus obtaining better results about word PMI-score.

\subsection{Topic Content Analysis}
\subsubsection{Topic Word Difference across Multiple Social Networks}
Perplexity is utilized to determine the topic number in related work\cite{heinrich2005parameter}. When the number of topics reaches 80, the perplexity progressively becomes smooth, as mentioned in the previous section. Therefore, we use the experiment results when the number of topics is 80 to show the general topic content of each network. The number of topics is also the same in work\cite{lee2017analyzing}.

The general topic words are shown in in Tab.\ref{tab_general_topic_content}.
We can find that the general topic contents that users like to discuss on Twitter are daily stuff, media events, etc. Twitter users use more oral words. There are larger verbs used by users compared with other social media platforms. Due to users' daily images on Instagram, classified as a photo-based social network, the objects in photos are typically people or scenes, and the subject is inclined to landscape, fashion, food, etc. Tumblr resembles a typical blog in appearance. Tumblr features more text material than Twitter, as well as more nouns and fewer vocal verbs. On the Tumblr network, users create content by combining words with other multimedia elements, such as photographs, to describe topics. The topic content shown in Tab.\ref{tab_general_topic_content} is about vehicles and business marketing, respectively.
\begin{table}[htb]
\caption{General Topic Contents}
\begin{center}
\begin{tabular}{ c|c|c|c|c|c }
\hline
\multicolumn{2}{ c|}{\textbf{Twitter}}& \multicolumn{2}{|c|}{\textbf{Instagram}}& \multicolumn{2}{|c }{\textbf{Tumblr}} \\
\hline
\textbf{Topic 1}&\textbf{Topic 2}& \textbf{Topic 1}&\textbf{Topic 2} &\textbf{Topic 1}&\textbf{Topic 2} \\
\hline
get&trump&food&sky&light&store\\
\hline
make&say&dish&tree&lead& kindle\\
\hline
people&president&cuisine&landscape&truck&bestseller\\
\hline
time&people&ingredient&nature&rear&new\\
\hline
know & white& food & sky&light& store\\
\hline
go & get& recipe & mountain&trailer& amazon\\
\hline
use & vote&meal&wilderness&lamp&buy\\
\hline
say & call&breakfast&environment&car&list\\
\hline
take & woman&lunch&plant&pair&author\\
\hline
work & make&meat&cloud&indicator&visit\\
\hline
\end{tabular}
\label{tab_general_topic_content}
\end{center}
\end{table}

Then we compare global and local topic content in each social network and understand users' latent motivations for using social networks.  We discover that, for one type of topic content, global topic content has an explicit relationship with each social network's local topic content. Different social network service providers can share data and give more comprehensive services to users by utilizing global topics and local topics.
As shown in Tab.\ref{tab_difference_topic_content1}, the topicA content is food, and the local topic has a large number of words that are explicitly embodied in the global topic. Global and local topics are more relevant. When a user prefers this type of topic, their preference tends to be consistent across networks.
\begin{table}[htb]
\caption{Difference of Topic Content across Multiple Social Networks 1}
\begin{center}
\begin{tabular}{ c|c|c|c }
\hline
\multicolumn{4}{c}{\textbf{TopicA}}\\
\hline
\textbf{Global}&\textbf{Local(Twitter)}& \textbf{Local(Instagram)}&\textbf{Local(Tumblr)}\\
\hline
food&food&food&food\\
\hline
cuisine&coffee&cuisine&chocolate\\
\hline
dish&new&dish&good\\
\hline
dessert&restaurant&dessert&make\\
\hline
ingredient&good&ingredient&day\\
\hline
good&get&cake&get\\
\hline
cake&eat&cream&recipe\\
\hline
cream&day&good&new\\
\hline
chocolate&best&meal&restaurant\\
\hline
coffee&make&chocolate&chocolate\\
\hline
\end{tabular}
\label{tab_difference_topic_content1}
\end{center}
\end{table}
The other type of topic content, shown in Tab.\ref{tab_difference_topic_content2}, is that local topic content is the implicit correlation to the global topic. The notable words in the local topic content are not the same as keywords in the global topic content. TopicB is a topic about residential design, local topic content in Instagram reflects ``plant'' decorations, and local topic content in Tumblr focuses on the room design itself.
The local topic reflects the influence of social networks on users' topic content. For latent semantic content, some things described in global topic content can be examined and understood through local topic content. It is feasible to deliver customized demand services to users in each social network based on this type of topic.
\begin{table}[htb]
\caption{Difference of Topic Content across Multiple Social Networks 2}
\begin{center}
\begin{tabular}{ c|c|c|c}
\hline
\multicolumn{4}{c}{\textbf{TopicB}} \\
\hline
\textbf{Global}&\textbf{Local(Twitter)}& \textbf{Local(Instagram)}&\textbf{Local(Tumblr)} \\
\hline
plant&wakafire&plant&room\\
\hline
flower&image&flower&design\\
\hline
room&garden&tree&furniture\\
\hline
tree&flower&petal&building\\
\hline
design&new&botany&house\\
\hline
furniture&post&leaf&property\\
\hline
house&poem&spring&architecture\\
\hline
botany&day&family&home\\
\hline
petal&location&pink&table\\
\hline
building&plant&garden&floor\\
\hline
\end{tabular}
\label{tab_difference_topic_content2}
\end{center}
\end{table}

\subsubsection{Word Distribution over Topic across Multiple Social Networks}
We can quantitatively explore the similarities and differences between global topics and local topics across multiple social networks.

Based on the probability distribution of word over topic $p(w|z)$, i.e., $\varphi_z^p$ or $\varphi_z^s$ in this paper, we can calculate the similarity between two word distributions over topic across social networks. This paper uses Jensen-Shannon Divergence (JSD) to measure the similarity, which takes values in the range of $[0,1]$, and two probability distributions are more similar to each other with respect to the lower JSD score.

\begin{figure}[htb]
\centerline{\includegraphics[width=0.4\textwidth]{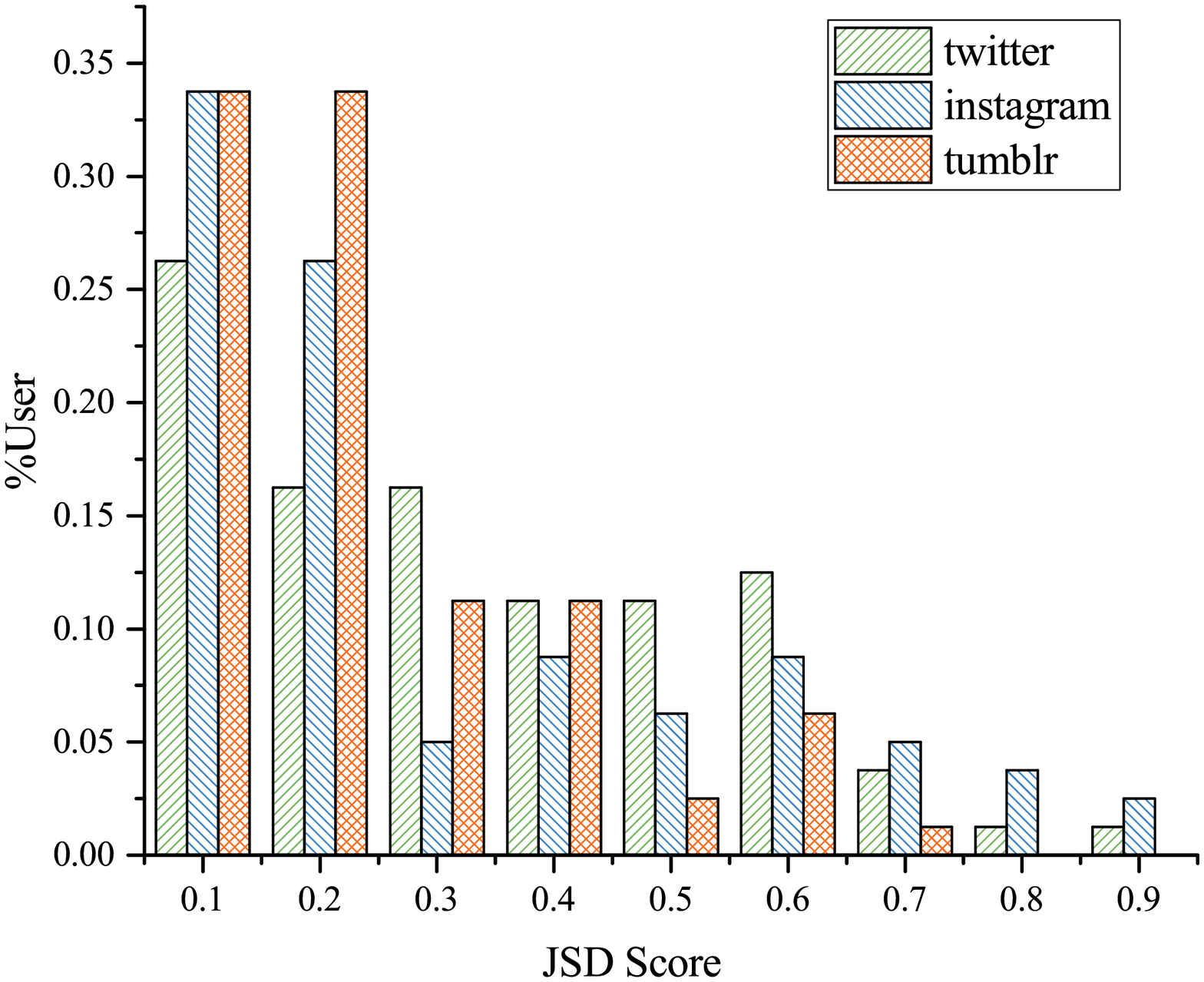}}
\caption{The JSD Score Distribution over Identical Local Topic}
\label{fig_JSD1}
\end{figure}

We first calculate JSD scores between each identical local topic across social networks. Then we can obtain proportion distribution based on JSD scores. The JSD score proportion distribution of identical local topics across social networks is shown in Fig.\ref{fig_JSD1}. 
Overall, the JSD values are concentrated in the middle of distribution in most cases, reflecting the consistency and complementarity are co-exist across multiple social networks as described in this paper.

\begin{figure}[htb]
\centerline{\includegraphics[width=0.4\textwidth]{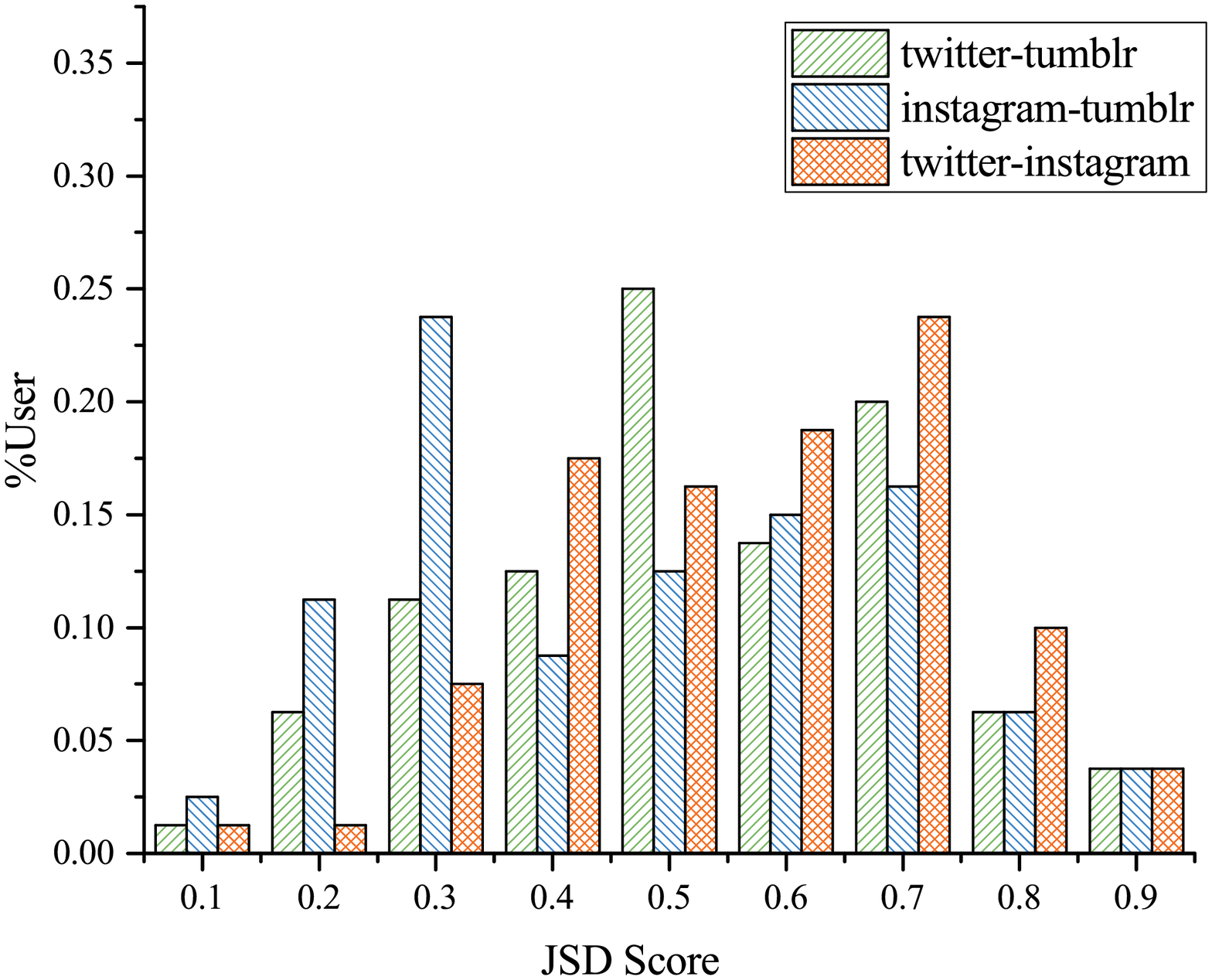}}
\caption{The JSD Score Distribution between Local Topic and Global Topic}
\label{fig_JSD2}
\end{figure}

Similarly, we can also obtain the JSD similarity between the local topic and the global topic in each social network, i.e., the JSD score between $\varphi_z^s$ and $\varphi_z^p$. The JSD score proportion distribution between local topic and global topic across social networks is shown in Fig.\ref{fig_JSD2}. 
It can be seen that most of the distributed JSD values are concentrated in the interval less than 0.3, which indicates the consistency of topic content between global topic and local topics.
\section{Conclusion}
In this paper, we propose a user topic preference model for multiple social networks. The model integrates a variety of data in different social networks to mine the topic preferences of users in multiple social networks. Experiments show that the model described in this research outperforms prior works and is capable of effectively mining the differences in users' topic content across social networks. However, the topic contents reported in the experiment section still have noise that is difficult to decipher, which could be due to the sparse and colloquial usage of user content in social networks. Better data preprocessing could help solve the situation. In the future, we'll look into more fine-grained methods for inferring users' personalized topic preferences, more diverse multimedia data to model user preferences, and an evolution model of user topic preferences over time.
\section*{Acknowledgment}
This work is supported by Natural Science Foundation of China under Grants No. 61772133, No.61972087.National Social Science Foundation of China under Grants No. 19@ZH014. Jiangsu Provincial Key Project under Grants No.BE2018706.Natural Science Foundation of Jiangsu province under Grants No.SBK2019022870. Jiangsu Provincial Key Laboratory of Computer Networking Technology. Jiangsu Provincial Key Laboratory of Network and Information Security under Grants No. BM2003201, and Key Laboratory of Computer Network and Information Integration of Ministry of Education of China under Grants No. 93K-9.

\bibliographystyle{IEEEtran}
\bibliography{IEEEabrv,mybibfile}

\begin{thebibliography}{10}
\providecommand{\url}[1]{#1}
\csname url@samestyle\endcsname
\providecommand{\newblock}{\relax}
\providecommand{\bibinfo}[2]{#2}
\providecommand{\BIBentrySTDinterwordspacing}{\spaceskip=0pt\relax}
\providecommand{\BIBentryALTinterwordstretchfactor}{4}
\providecommand{\BIBentryALTinterwordspacing}{\spaceskip=\fontdimen2\font plus
\BIBentryALTinterwordstretchfactor\fontdimen3\font minus
  \fontdimen4\font\relax}
\providecommand{\BIBforeignlanguage}[2]{{%
\expandafter\ifx\csname l@#1\endcsname\relax
\typeout{** WARNING: IEEEtran.bst: No hyphenation pattern has been}%
\typeout{** loaded for the language `#1'. Using the pattern for}%
\typeout{** the default language instead.}%
\else
\language=\csname l@#1\endcsname
\fi
#2}}
\providecommand{\BIBdecl}{\relax}
\BIBdecl

\bibitem{zhou2015cross}
X.~Zhou, X.~Liang, H.~Zhang, and Y.~Ma, ``Cross-platform identification of
  anonymous identical users in multiple social media networks,'' \emph{IEEE
  transactions on knowledge and data engineering}, vol.~28, no.~2, pp.
  411--424, 2015.

\bibitem{zhang2014meta}
J.~Zhang, P.~S. Yu, and Z.-H. Zhou, ``Meta-path based multi-network collective
  link prediction,'' in \emph{Proceedings of the 20th ACM SIGKDD international
  conference on Knowledge discovery and data mining}, 2014, pp. 1286--1295.

\bibitem{cao2016joint}
X.~Cao and Y.~Yu, ``Joint user modeling across aligned heterogeneous sites,''
  in \emph{Proceedings of the 10th ACM Conference on Recommender Systems},
  2016, pp. 83--90.

\bibitem{li2016influence}
G.-L. Li, Y.-P. Chu, J.-H. Feng, and Y.-Q. XU, ``Influence maximization on
  multiple social networks [j],'' \emph{Chinese Journal of Computers}, vol.~39,
  no.~4, pp. 643--656, 2016.

\bibitem{zhang2015least}
H.~Zhang, D.~T. Nguyen, H.~Zhang, and M.~T. Thai, ``Least cost influence
  maximization across multiple social networks,'' \emph{IEEE/ACM Transactions
  on Networking}, vol.~24, no.~2, pp. 929--939, 2015.

\bibitem{philip2015mcd}
S.~Y. Philip and J.~Zhang, ``Mcd: mutual clustering across multiple social
  networks,'' in \emph{2015 IEEE International Congress on Big Data}.\hskip 1em
  plus 0.5em minus 0.4em\relax IEEE, 2015, pp. 762--771.

\bibitem{zhu2019community}
Z.~Zhu, T.~Zhou, C.~Jia, W.~Liu, B.~Liu, and J.~Cao, ``Community detection
  across multiple social networks based on overlapping users,''
  \emph{Transactions on Emerging Telecommunications Technologies}, p. e3928,
  2019.

\bibitem{lee2017analyzing}
R.~K.-W. Lee, T.-A. Hoang, and E.-P. Lim, ``On analyzing user topic-specific
  platform preferences across multiple social media sites,'' in
  \emph{Proceedings of the 26th International Conference on World Wide Web},
  2017, pp. 1351--1359.

\bibitem{boyd2007social}
D.~M. Boyd and N.~B. Ellison, ``Social network sites: Definition, history, and
  scholarship,'' \emph{Journal of computer-mediated Communication}, vol.~13,
  no.~1, pp. 210--230, 2007.

\bibitem{hong2010empirical}
L.~Hong and B.~D. Davison, ``Empirical study of topic modeling in twitter,'' in
  \emph{Proceedings of the first workshop on social media analytics}, 2010, pp.
  80--88.

\bibitem{cao2017you}
C.~Cao, H.~Ge, H.~Lu, X.~Hu, and J.~Caverlee, ``What are you known for?
  learning user topical profiles with implicit and explicit footprints,'' in
  \emph{Proceedings of the 40th International ACM SIGIR Conference on Research
  and Development in Information Retrieval}, 2017, pp. 743--752.

\bibitem{chen2017discerning}
E.~Chen, G.~Zeng, P.~Luo, H.~Zhu, J.~Tian, and H.~Xiong, ``Discerning
  individual interests and shared interests for social user profiling,''
  \emph{World Wide Web}, vol.~20, no.~2, pp. 417--435, 2017.

\bibitem{you2016picture}
Q.~You, S.~Bhatia, and J.~Luo, ``A picture tells a thousand words—about you!
  user interest profiling from user generated visual content,'' \emph{Signal
  Processing}, vol. 124, pp. 45--53, 2016.

\bibitem{farnadi2018user}
G.~Farnadi, J.~Tang, M.~De~Cock, and M.-F. Moens, ``User profiling through deep
  multimodal fusion,'' in \emph{Proceedings of the Eleventh ACM International
  Conference on Web Search and Data Mining}, 2018, pp. 171--179.

\bibitem{mislove2010you}
A.~Mislove, B.~Viswanath, K.~P. Gummadi, and P.~Druschel, ``You are who you
  know: inferring user profiles in online social networks,'' in
  \emph{Proceedings of the third ACM international conference on Web search and
  data mining}, 2010, pp. 251--260.

\bibitem{chen2016profiling}
J.~Chen, J.~He, L.~Cai, and J.~Pan, ``Profiling online social network users via
  relationships and network characteristics,'' in \emph{2016 IEEE Global
  Communications Conference (GLOBECOM)}.\hskip 1em plus 0.5em minus 0.4em\relax
  IEEE, 2016, pp. 1--6.

\bibitem{xu2012modeling}
Z.~Xu, Y.~Zhang, Y.~Wu, and Q.~Yang, ``Modeling user posting behavior on social
  media,'' in \emph{Proceedings of the 35th international ACM SIGIR conference
  on Research and development in information retrieval}, 2012, pp. 545--554.

\bibitem{hu2015community}
Z.~Hu, J.~Yao, B.~Cui, and E.~Xing, ``Community level diffusion extraction,''
  in \emph{Proceedings of the 2015 ACM SIGMOD International Conference on
  Management of Data}, 2015, pp. 1555--1569.

\bibitem{feng2019user}
J.~Feng, Y.~Rao, H.~Xie, F.~L. Wang, and Q.~Li, ``User group based emotion
  detection and topic discovery over short text,'' \emph{World Wide Web}, pp.
  1--35, 2019.

\bibitem{wang2019community}
Y.~Wang, D.~Jin, K.~Musial, and J.~Dang, ``Community detection in social
  networks considering topic correlations,'' in \emph{Proceedings of the AAAI
  Conference on Artificial Intelligence}, vol.~33, no.~01, 2019, pp. 321--328.

\bibitem{abel2011analyzing}
F.~Abel, S.~Ara{\'u}jo, Q.~Gao, and G.-J. Houben, ``Analyzing cross-system user
  modeling on the social web,'' in \emph{International Conference on Web
  Engineering}.\hskip 1em plus 0.5em minus 0.4em\relax Springer, 2011, pp.
  28--43.

\bibitem{abel2013cross}
F.~Abel, E.~Herder, G.-J. Houben, N.~Henze, and D.~Krause, ``Cross-system user
  modeling and personalization on the social web,'' \emph{User Modeling and
  User-Adapted Interaction}, vol.~23, no.~2, pp. 169--209, 2013.

\bibitem{zhao2011comparing}
W.~X. Zhao, J.~Jiang, J.~Weng, J.~He, E.-P. Lim, H.~Yan, and X.~Li, ``Comparing
  twitter and traditional media using topic models,'' in \emph{European
  conference on information retrieval}.\hskip 1em plus 0.5em minus 0.4em\relax
  Springer, 2011, pp. 338--349.

\bibitem{qian2016multi}
S.~Qian, T.~Zhang, and C.~Xu, ``Multi-modal multi-view topic-opinion mining for
  social event analysis,'' in \emph{Proceedings of the 24th ACM international
  conference on Multimedia}, 2016, pp. 2--11.

\bibitem{heinrich2005parameter}
G.~Heinrich, ``Parameter estimation for text analysis,'' Technical report,
  Tech. Rep., 2005.

\bibitem{cheng2014btm}
X.~Cheng, X.~Yan, Y.~Lan, and J.~Guo, ``Btm: Topic modeling over short texts,''
  \emph{IEEE Transactions on Knowledge and Data Engineering}, vol.~26, no.~12,
  pp. 2928--2941, 2014.

\end{thebibliography}

\end{document}